\documentclass[aps,prb,twocolumn,showpacs,longbibliography,superscriptaddress,final,floatfix]{revtex4-1}
\usepackage{amsmath}
\usepackage{amsfonts}
\usepackage{amssymb}
\usepackage{physics}
\usepackage{color}
\usepackage{hyperref}
\usepackage[]{inputenc}
\usepackage{enumerate}
\usepackage{graphicx}
\usepackage{float}
\usepackage{amsmath}
\usepackage{amsfonts}
\usepackage{lipsum}
\usepackage{dcolumn}
\usepackage{hyperref}
\usepackage{subfigure}
\usepackage{epsfig}
\usepackage{epsf} 
\usepackage{epstopdf}

\usepackage{color}
\usepackage{enumitem}

\usepackage{xr}
\makeatletter

\begin{document}
	
	
	\title{ Super-diffusive transport in two-dimensional Fermionic wires }
	
	
	\author{Junaid Majeed Bhat\\\textit{Department of Physics, Faculty of Mathematics and Physics, \\University of Ljubljana, 1000 Ljubljana, Slovenia}}
	

	
	
	\date{\today}
	
	\begin{abstract}
	We present a two-dimensional model of a Fermionic wire which shows a power-law conductance behavior despite  the presence of uncorrelated disorder along the direction of the transport. The power-law behavior  is attributed to the presence of energy eigenstates of diverging localization length  below some energy cutoff, $E_c$. To study transport, we place the wire in contact with electron reservoirs biased around a Fermi level, $E$. We show that the conductance scales super-diffusively for $|E|<E_c$ and decays exponentially for $|E|>E_c$. At $|E|=E_c$, we show that the conductance scales diffusively or with different sub-diffusive power-laws depending  on the sign of the expectation value of the disorder and the parameters of the wire.

	\end{abstract}
	
	\maketitle
Introduction of disorder in quadratic  systems is known to generically cause localization  since the seminal work of Anderson~\cite{PhysRev.109.1492} and other subsequent works~\cite{10.1143/PTP.45.1713,PhysRevLett.42.673,RevModPhys.57.287,KRISHNA2021168537}. Thus, the conductance decays exponentially with the system size in disordered systems. On the other hand, the conductance in disorder free quadratic  systems  is ballistic. While these  behaviors of the conductance are the two extremes of perfectly conducting and insulating behaviors, respectively, conductance can also scale as a power-law$~(1/N^{\alpha})$ with the system size, $N$. The scaling can be categorized into diffusive $(\alpha=1)$, super-diffusive $(\alpha<1)$ and sub-diffusive $(\alpha>1)$ behaviors. Diffusive scaling is in general expected on addition of interactions to quadratic systems~\cite{RevModPhys.93.025003}. Super-diffusive and sub-diffusive are therefore considered anomalous. Sub-diffusive transport has been observed in disordered interacting spin systems~\cite{PhysRevLett.117.040601,de2020subdiffusion,PhysRevB.93.224205,PhysRevB.107.014207} and also at the band edges of quadratic Fermionic systems~\cite{PhysRevLett.130.187101,PhysRevB.109.125415,RaySub}. Super-diffusive transport is observed in quadratic systems by introducing certain types of correlated~\cite{PhysRevLett.65.88,wang2024superdiffusive} or aperiodic~\cite{doi:10.1143/JPSJ.57.230} disorders.  Recent studies have shown super-diffusive transport in interacting spin systems~\cite{ljubotina2017spin,PhysRevLett.122.210602,PhysRevLett.106.220601} and also in certain type of dephasing models~\cite{wang2023superdiffusive}.

	The phenomenon of Anderson localization also occurs in classical systems of   harmonic wires~\cite{10.1143/PTPS.45.56,PhysRevLett.58.2486,PhysRevB.27.5592}. Nevertheless, certain peculiar  systems are known to show power-law scaling of energy current even in the presence of uncorrelated  disorder~\cite{10.1063/1.1665794,10.1063/1.1665793,Cane_2021}. A simple example is a 1D harmonic wire with disordered masses first studied in detail  by Casher and Lebowitz~\cite{10.1063/1.1665794}. In this case, the localization length of the normal modes of the wire diverges as the conducting frequency approaches zero~\cite{10.1143/PTPS.45.56}.   Therefore, low frequency modes contribute to the transport  effectively giving rise to a power-law scaling of the energy current. The exact scaling is determined by the behavior of the localization length as the frequency, $\omega\rightarrow 0$ and the low-frequency behavior of the heat transmission~\cite{verheggen1979transmission,PhysRevLett.86.5882,PhysRevE.78.051112,ajanki2011rigorous,de2017step,Cane_2021}.   
	
	To our knowledge, such peculiar models have not been discussed in the context of Fermionic wires. In this work, we introduce a simple 2D model for  Fermionic wires   that realizes  physics analogous to 1D  mass disordered harmonic wires.  Therefore, our model shows a power-law scaling of the conductance  in presence of uncorrelated disorder along the direction of the transport. This is  in contrast with the earlier studies of 1D Fermionic wires where the disorder is either correlated~\cite{PhysRevLett.65.88,wang2024superdiffusive} or aperiodic~\cite{doi:10.1143/JPSJ.57.230}. While in the harmonic wires the localization length diverges only at a particular point in the energy spectrum, our 2D model  contains eigenfunctions of diverging localization lengths at energies with absolute values less than some cut off, $E_c$. Therefore, $E_c$ effectively behaves as a mobility edge.


	
	To study electron transport, we consider a sample of the wire of size $L\times W$ and place it in contact with  metallic leads along   its vertically opposite edges. We then employ the non-equilibrium Green's function formalism (NEGF), to look at the scaling of the conductance with $L$  while the leads are biased around a Fermi-level, $E$. We find that for $|E|<E_c$, the conductance shows a super-diffusive behavior and for $|E|>E_c$ the conductance scales exponentially with the length of the wire. At the transition point $|E|=E_c$,  the conductance scales with  different power-laws for  positive, negative or zero  expectation value of the disorder. We present heuristic arguments  that explain all the numerically observed power-laws. However, the underlying assumption of these arguments fails for the cases  where $|E|=E_c$, and the expectation value of disorder vanishes. In these cases, the observed power-laws are underestimated by a factor of $L$ by the theoretical arguments, opening up an interesting problem for further studies.    
	
		This paper is structured as follows: In Sec.~(\ref{sec:model}), we lay down the details of the wire, leads, and the contacts between them. We also discuss the eigenfunctions of the isolated wire and illustrate why we expect diverging localization lengths  for certain range of energies. In Sec.~(\ref{sec:negf}), we set up the NEGF formalism which we use to look at the average behavior of the conductance. We then consider the Lyapunov exponents, inverse of the localization length, in Sec.~(\ref{sec:lyap}) for the wave-functions of the wire and present results for its asymptotic behavior around the point where it vanishes. In the penultimate section, Sec.~(\ref{sec:plaw}), we use the NEGF expression for the conductance and  the knowledge of the Lyapunov exponents to determine the different power-laws for the conductance and  also present numerical results for its behavior. We conclude in Sec.~(\ref{sec:concl}).


	\section{The model}
	\label{sec:model}
	We consider the wire Hamiltonian, $\mathcal{H}_W$, to be given by a tight-binding model defined on a rectangular lattice of size $L\times W$.  
	Let us label the  annihilation and creation operators, satisfying usual anti-commutation relations, at a site $(x,y)$ as ($\psi(x,y)$, $\psi^\dagger(x,y)$) on the wire. 
The Hamiltonian of the wire is then given by,
	\begin{align}
		\label{eq:wirehamil}\mathcal{H}_W&=\sum_{x=1}^{L}\epsilon_x \Psi^\dagger(x)H_0\Psi(x)+  \sum_{x=1}^{L-1}\left[\Psi^\dagger(x)\Psi(x+1)+ \text{h.c.}\right],
	\end{align}
where, $\Psi(x)=(\psi(x,1),\psi(x,2),..,\psi(x,W))^T$. $\epsilon_x$ is the disorder parameter chosen randomly at every $x$ with expectation value of $\expval{\epsilon}$.

The eigenstates of the wire are given by, $\psi_{E_{km}}(x,y)= \chi_k(y) \phi_{k,m}(x)$,  where $k=1,2,...,W$ and $m=1,2,3,...,L$.   $\chi_k$ is an eigenvector of $H_0$ with eigenvalue $\lambda_k$ and $\phi_{k,m}(x)$ satisfy,
\begin{align}
 \phi_{k,m}(x-1)+\lambda_k\epsilon_x\phi_{k,m}(x)+\phi_{k,m}(x+1)=E_{k,m}\phi_{k,m}(x).
 \end{align}
  The solutions for $\phi_{k,m}(x)$ correspond to the wavefunctions  of a 1D Anderson insulator of length $L$ with onsite disorder of effective  strength proportional to $\lambda_k$.  Therefore, the localization length of the eigenvectors depends on $\lambda_k$ and diverges if $\lambda_k$ vanishes. Thus, if $\lambda_k$'s are banded and cross the value zero, then in the limit $W\rightarrow\infty$ there are extended eigenstates with energies of absolute value less than $E_c=2$.  So,  a  fraction of the eigenfunctions of the wire which correspond to $\lambda_k$ near zero  contribute to the transport giving rise to a  power-law scaling of the conductance. This power-law is determined by the behavior of the localization length near $\lambda_k=0$, which we discuss later. While this physics holds for any choice of $H_0$, we now fix, for the rest of the paper, a simple choice of $H_0$ corresponding to nearest neighbor hopping with an onsite chemical potential of $\mu$, so that its spectrum is given by $\lambda_k=\mu+2\cos\frac{k\pi}{W+1}$.  
  \begin{figure}
 	\centering
 	\subfigure[$\expval{\epsilon}<0$]{
 		\includegraphics[width=0.15\textwidth,page=1]{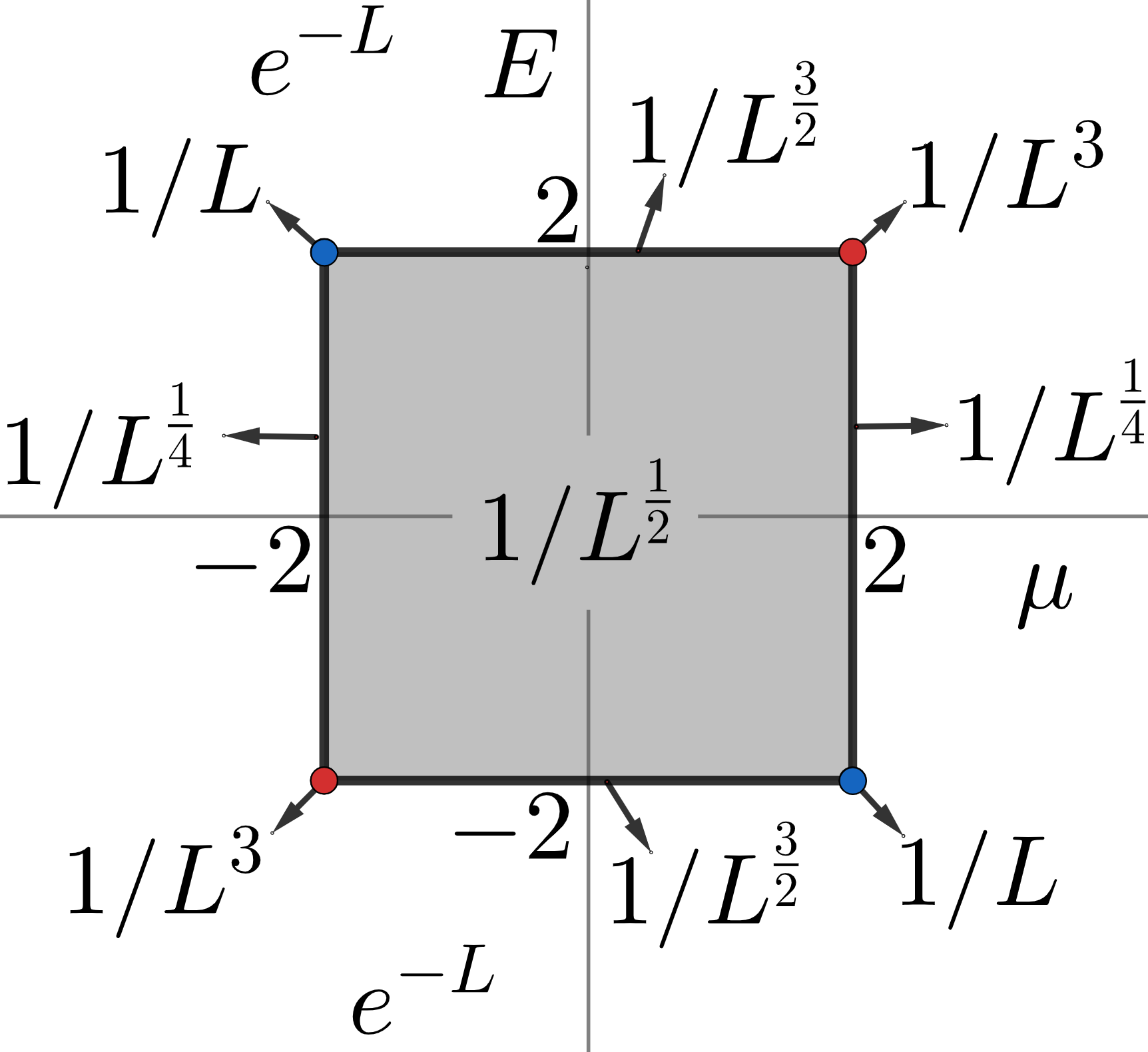}}
 	\subfigure[$\expval{\epsilon}=0$]{
 		\includegraphics[width=0.15\textwidth,page=2]{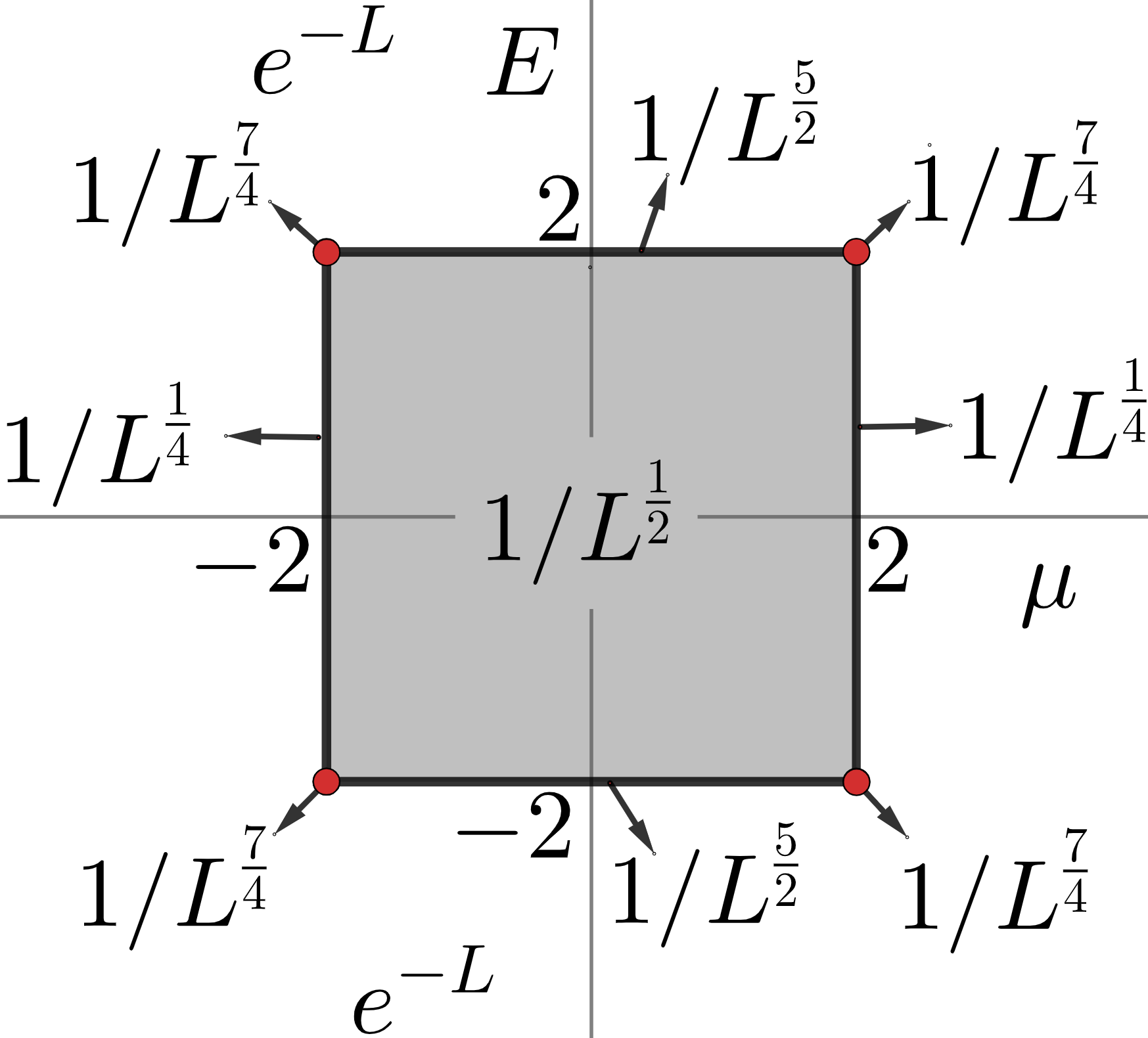}}
 	\subfigure[$\expval{\epsilon}>0$]{
 		\includegraphics[width=0.15\textwidth,page=2]{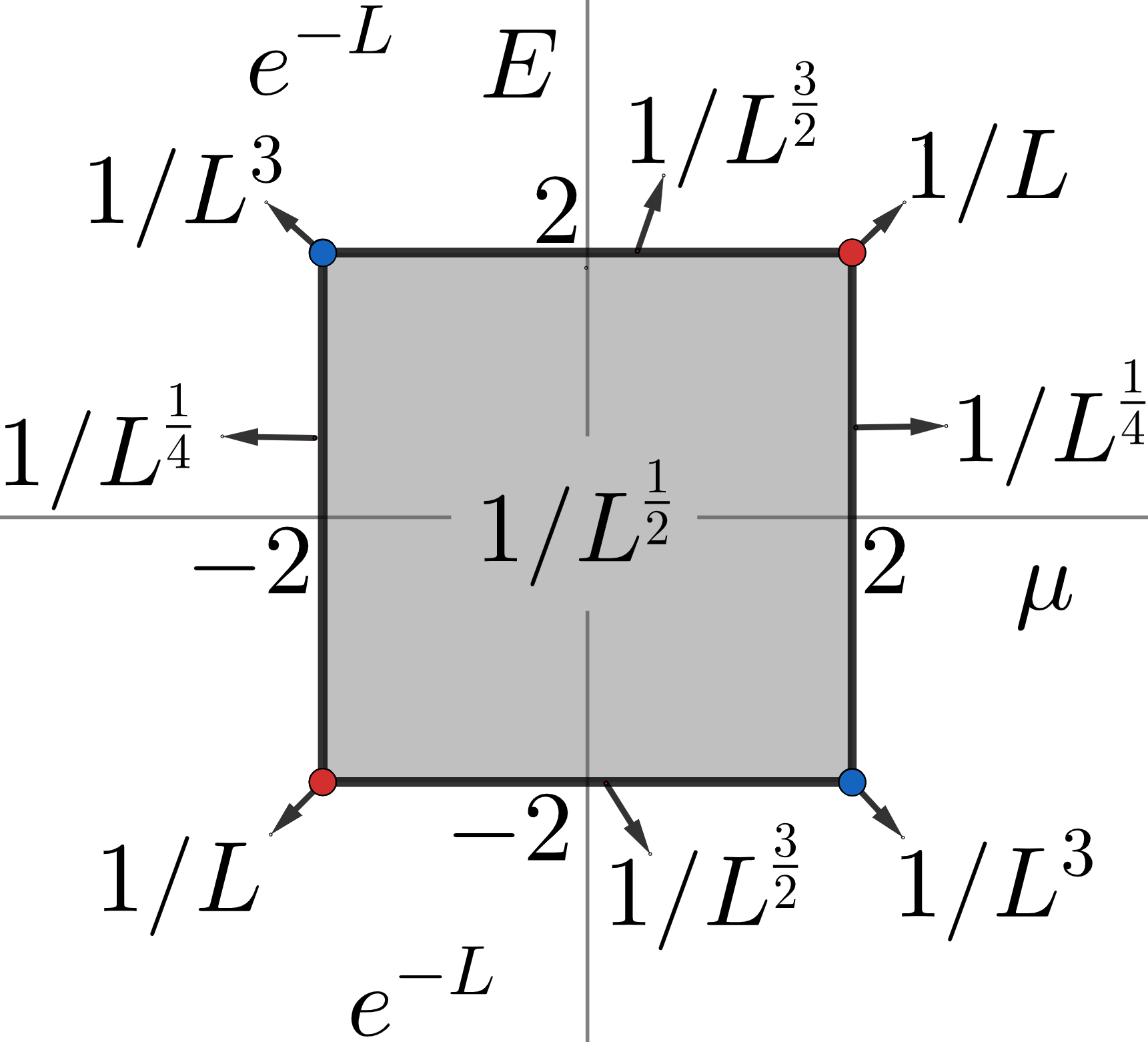}}
 	\vspace{-1.7\baselineskip}
 	\caption{Power laws for the behavior of the conductance at different Fermi Levels and the parameter $\mu$. }
 	\label{fig:chart}
 \end{figure}  

  To probe the electron transport across the wire, we place it in contact with metallic leads acting as electron reservoirs at its two opposite edges along $y-$direction. We then use the NEGF formalism to study the conductance  with the reservoirs kept at zero temperatures and biased around a Fermi-level, $E$. The two reservoirs  are themselves modeled as  nearest-neighbor tight-binding Hamitonians on a square lattice with hoppings along $x$ and $y$ directions  given by $\eta_{bx}$ and $\eta_{by}$, respectively.  The contacts  are also modeled as tight-binding Hamiltonians with hopping strength   $\eta_c$. For details of the reservoir and contact Hamiltonians see  Appendix \ref{app:res_hamil}.
  
  We obtain different behaviors of the conductance with respect to $E$ and $\mu$ that are summarized in Fig.~(\ref{fig:chart}).  Inside the shaded square and along its vertical edges in Fig.~(\ref{fig:chart}), we see a super-diffusive scaling of the conductance. Along the horizontal edges of the square and at the corners the conductance scales sub-diffusively except for some cases where it scales  diffusively, and shows an interesting dependence on the sign of the expectation value of the disorder. Outside the shaded square, the conductance scales as $e^{-L}$. The  behavior of the conductance is determined by the behavior of electron transmission at the Fermi level $E$ and the localization length near $\lambda_k=0$. So, let us  discuss the two separately.

	\section{NEGF Conductance}
	\label{sec:negf}
	The non-equilibrium steady state (NESS) of the wire can be obtained using the NEGF formalism 
 in terms of the effective non-equilibrium Green's function for the wire defined as~\cite{dhar2006, Bhat2020transport},
	$
		G^+(E)=(E-H_W-\Sigma_L(E)-\Sigma_R(E))^{-1},
	$
	where $\Sigma_L(E)$ and $\Sigma_R(E)$ are the self energy contributions due to the reservoirs, and $H_W$ is the full hopping matrix of the wire.
For the reservoirs kept at zero temperatures and their chemical potential  biased around a Fermi-level $E$, the conductance of the wire, in units of $e^2/h=1$, is given in terms of $G^+(E)$ as,
$
	T(E)=4\pi^2\Tr[G^+(E)\Gamma_R(E)G^-(E)\Gamma_L(E)],\label{eq:cond1}
$
where $G^-(E)=[G^+(E)]^\dagger$ and $\Gamma_{L/R}=(\Sigma_{L/R}^\dagger-\Sigma_{L/R})/(2\pi i)$. 

 While the above formula holds for arbitrary lattice models, for our model this could be approximated by
 \begin{align}
 	&\tau(E)=\frac{T(E)}{W}\approx\frac{4}{\pi}\int_0^\pi d\mathbf k |F_L(\lambda_\mathbf k)|^2,\label{eq:condint1}
 \end{align}
where $F_L(\lambda_\mathbf k)=\gamma^2/[p_L(\lambda_\mathbf k)+i \gamma [p_{L-1}(\lambda_\mathbf k)+q_{L}(\lambda_\mathbf k)] -\gamma^2 q_{L-1}(\lambda_\mathbf k)]$ and $\gamma=\frac{\eta_c^2}{\eta_{bx}}$, in the limit $W\rightarrow\infty$ and $|\eta_{bx}|>>|E|+|2\eta_{by}|$, see Appendix \ref{app:cond1}. The integral runs over the spectrum of $H_0$ in the limit $W\rightarrow\infty$, given by $\mu+2\cos(\mathbf{k}),~\mathbf{k}\in(0,\pi)$.
  $p_L(\lambda_\mathbf k)$ and $q_L(\lambda_\mathbf k)$  are obtained via same iteration equations, which for $p_L(\lambda_\mathbf k)$ reads
$
	p_{i+1}(\lambda_\mathbf k)=(-E+\epsilon_{i+1}\lambda_\mathbf k)p_i(\lambda_\mathbf k)-p_{i-1}(\lambda_\mathbf k),\label{eq:pi}
$
but with different   initial conditions. Therefore, it suffices to consider only one of the two. It is clear that the asymptotic behavior of the  $p_L(\lambda_\mathbf k)$ with $L$  eventually controls the behavior of the conductance with $L$. Also, note that the iteration equation for $p_i(\lambda_\mathbf k)$ is the same as the equation for $\phi_{k,m}(i)$. Therefore, the asymptotic behavior of $p_L(\lambda_\mathbf k)$ also gives the localization length of the eigenvectors of the wire. We now discuss the behavior of $p_L(\lambda_\mathbf k)$ as $\lambda_\mathbf k\rightarrow 0$, which we later use to determine the scaling of the conductance.

\begin{widetext}
	
	\begin{figure}[h!]	
	
		\centering
		\subfigure{
			\includegraphics[width=0.23\textwidth,page=1]{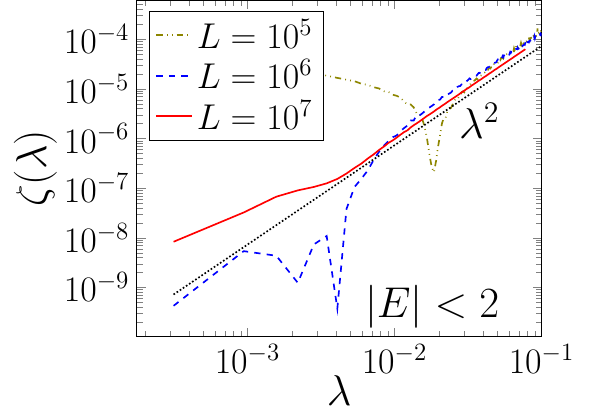}}
		\subfigure{
			\includegraphics[width=0.23\textwidth,page=2]{plot_lyap.pdf}}
		\subfigure{
			\includegraphics[width=0.23\textwidth,page=3]{plot_lyap.pdf}}
		\subfigure{
			\includegraphics[width=0.23\textwidth,page=4]{plot_lyap.pdf}}
		\vspace{-1.5\baselineskip}
		\caption{ Lyapunov exponents for the iteration equation, Eq.~(\ref{eq:itereq})  for $|E|<2$ and $E=2$ with positive, negative and zero expectation value of the disorder. For $E=-2$, $\zeta(\lambda)\sim \lambda$ for $\expval{\epsilon}<0$ and $\zeta(\lambda)\sim \lambda^{1/2}$ for $\expval{\epsilon}>0$. The data presented is averaged over $10^3$ realizations of the disorder chosen uniformly from the intervals $(-1,0)$, $(-1,1)$ and $(0,1)$ for negative, zero and positive expectation value cases, respectively.}
	
		\label{fig:lyap}
	\end{figure}  
\end{widetext}
\section{Lyapunov Exponents}
	\label{sec:lyap}
	Let us consider the iteration equation for $p_L(\lambda)$, we have
	\begin{equation}
		p_{i+1}(\lambda)=(-E+\epsilon_{i+1}\lambda)p_i(\lambda)-p_{i-1}(\lambda).\label{eq:itereq}
	\end{equation}
We have dropped the $\mathbf k$ subscript on $\lambda$ for now, and we will only consider $\lambda>0$, as $\lambda<0$ is equivalent to shifting the sign of the disorder parameter. 
 A theorem due to  Furstenberg~\cite{furstenberg1963noncommuting},  guarantees non-negativity and the existence of the Lyapunov exponent, inverse of the localization length, defined as,
\begin{align}
 	\zeta(\lambda)=\lim\limits_{L\rightarrow\infty} \frac{1}{L}\expval{\log|p_L(\lambda)|},
\end{align}
for any initial condition, and therefore $p_L(\lambda)\sim e^{L\zeta(\lambda)}$. 

For $|E|\leq2$, $\zeta(\lambda)\rightarrow 0$ as $\lambda\rightarrow 0$.
Therefore, near zero $\lambda$ values such that $L\zeta(\lambda)<1$, contribute in Eq.~(\ref{eq:condint1}).   Thus, the behavior of $\zeta(\lambda)$ near $\lambda=0$ is  crucial to the scaling of the conductance. We present numerical results on this behavior at different $E$ in Fig.~(\ref{fig:lyap}). We see that while for $|E|<2$, $\zeta(\lambda)\sim \lambda^2$ irrespective of the expectation value of the disorder, for $|E|=2$ the behavior of $\zeta(\lambda)$ is different for different signs of the expectation value of the disorder. For $E=2$, we find  $\zeta(\lambda)\sim\lambda$, $\zeta(\lambda)\sim\lambda^{2/3}$, and $\zeta(\lambda)\sim\lambda^{1/2}$ for positive, zero and negative expectation value of the disorder, respectively. For $E=-2$, the behaviors are the same as $E=2$ except that $\zeta(\lambda)\sim \lambda^{1/2}$ for positive expectation value and $\zeta(\lambda)\sim \lambda$ for negative expectation value of the disorder.

We outline the proof here and present the details  in the Appendix \ref{appen:lyap}. Let us consider the case of  $|E|<2$. We follow the steps of Matsuda and Ishii in Ref.~\onlinecite{10.1143/PTPS.45.56} as they considered the iteration equation, Eq.~(\ref{eq:itereq}), for $E=-2$ and $\expval{\epsilon}<0$  in the context of classical harmonic wires with mass disorder.  We start by making the change of variables, \begin{equation}\frac{p_{n+1}(\lambda)}{p_{n}(\lambda)}=\frac{\cos(\theta_n+h(\lambda,E))}{\cos\theta_n},\end{equation} where $2\cos h(\lambda,E)= -E+\expval{\epsilon}\lambda$ so that the iteration equation now becomes,
\begin{equation}
	\label{eq:itertheta}
	\theta_{n+1}=\arctan\left[\tan(\theta_n+h(\lambda,E))+\frac{\lambda(\epsilon-\expval{\epsilon})}{\sin h(\lambda,E)}\right]=\bar \Theta[\theta_n,\lambda].
\end{equation}
The Lyapunov exponent in terms of the variable $\theta$, is given by
\begin{align}
	\zeta(\lambda)
	&=\lim\limits_{L\rightarrow\infty}\frac{1}{L} \expval{\sum_{n=1}^L\log\abs{ \frac{\cos(\theta_n+h(\lambda,E))}{\cos\theta_n}}},
	\\&=\int_{-\pi/2}^{\pi/2}d\theta ~\mathbf P[\theta,\lambda] \log\abs{ \frac{\cos(\theta+h(\lambda,E))}{\cos\theta}}\label{eq:zetatheta}.
\end{align}
In the last step we have replaced the average over "time" in the Marko process defined by the iteration equation, Eq.~(\ref{eq:itereq}), by the integral over the invariant distribution $\mathbf P[\theta,\lambda]$ of the Marko process in accordance with the ergodic hypothesis.  Expanding Eq.~(\ref{eq:zetatheta}) in orders of $\lambda$ we have,
\begin{align}
	\zeta(\lambda)\notag= \int_{-\pi/2}^{\pi/2}&d\theta\log|\cos\theta|\bigg[\boldsymbol{\zeta}_0(\theta,E)+ \lambda\boldsymbol{\zeta}_1(\theta,E)\notag\\&+\lambda^2\boldsymbol{\zeta}_2(\theta,E)+\order{\lambda^3}\bigg].
\end{align}
Assuming $h(\lambda,E)=\sum_{i=0}^{\infty} h_i(E) \lambda^i$ and $\mathbf{P}[\theta,\lambda]=\sum_{i=0}^\infty\mathbf{P}_i(\theta) \lambda^{i}$, the coefficients $\boldsymbol{\zeta}_i(\theta,E)$  depend on $\theta$ and $E$ via the functions $\mathbf P_i(\theta)$ and $h_i(E)$. Therefore, we need to determine $\mathbf P[\theta,\lambda]$ at least up to   order  $\lambda^2$ to find  $\zeta(\lambda)$. It can be shown that the invariant distribution satisfies the self-consistent equation,
\begin{equation}
	\mathbf{P}[\theta,\lambda]-\int d\epsilon~\mathbf P[\Theta[\theta,\lambda],\lambda] \partial_\theta\Theta[\theta,\lambda]\mathbf{p}(\epsilon)=0,\label{eq:selfP}
\end{equation}
where $\mathbf{p}(\epsilon_x)$ is the probability distribution for $\epsilon_x$ and
\begin{equation}
	\Theta[\theta,\lambda]=\arctan\left[\tan\theta-\frac{\lambda(\epsilon-\expval{\epsilon})}{\sin h(\lambda,E)}\right]-h(\lambda,E),
\end{equation}
is the inverse of the function $\bar\Theta[\theta,\lambda]$. Expanding the L.H.S of  Eq.~(\ref{eq:selfP}) in orders of $\lambda$ and setting each order to zero gives, $\boldsymbol{\zeta}_0(\theta,E)=0$, $\boldsymbol{\zeta}_1(\theta,E)=0$, and $\boldsymbol{\zeta}_2(\theta,E)\neq 0$ for the zeroth, first and second order in $\lambda$, respectively.
Hence, the  zeroth order and the first order terms vanish in the expansion of $\zeta(\lambda)$ and therefore  $\zeta(\lambda)\sim\lambda^2$ as $\lambda\rightarrow0$.
	\begin{widetext}
	
	\begin{figure}[h!]
		\centering
		\subfigure{
			\includegraphics[width=0.23\textwidth,page=1]{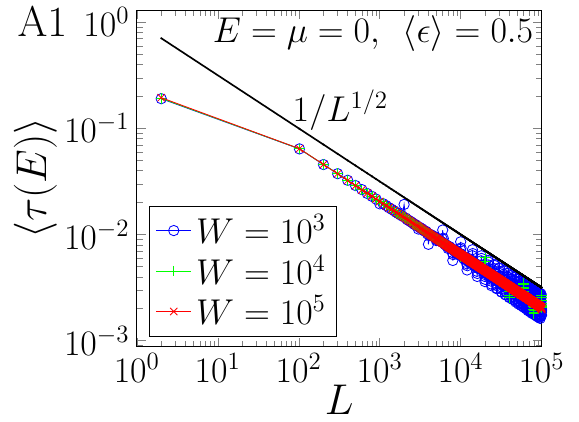}}
		\subfigure{
			\includegraphics[width=0.23\textwidth,page=4]{plot.pdf}}
		\subfigure{
			\includegraphics[width=0.23\textwidth,page=2]{plot.pdf}}
		\subfigure{
			\includegraphics[width=0.23\textwidth,page=3]{plot.pdf}}
		\subfigure{
			\includegraphics[width=0.23\textwidth,page=5]{plot.pdf}}
		\subfigure{
			\includegraphics[width=0.23\textwidth,page=6]{plot.pdf}}
		\subfigure{
			\includegraphics[width=0.23\textwidth,page=7]{plot.pdf}}
		\subfigure{
			\includegraphics[width=0.23\textwidth,page=8]{plot.pdf}}
		\vspace{-1.5\baselineskip}
		\caption{ Numerically observed power-laws for different Fermi levels and the parameter, $\mu$. We see a good agreement with the predicted values as the width of the wire is increased except for the cases B1 and B2. For these two cases the theoretical arguments predict $1/L^{7/2}$ and $1/L^{11/4}$, respectively. The theoretical arguments predict slower transport as these arguments underestimate  $\expval{|F_L(\lambda_\mathbf k)|^2}$  as can be seen from panel X. The data presented is averaged over $10^3$ disorder realizations. Parameter values for X, $W=10L=10^4$, $E=2$, $\mu=0$ and $\expval{\epsilon}=0$. }
	
		\label{fig:powlaw1}
	\end{figure}
\end{widetext}
	The proof relies on the fact that $\theta\in (-\pi/2,\pi/2)$ is bounded which is true as long as $h(\lambda,E)$ is real or equivalently $|-E+\expval{\epsilon}\lambda|<2$. Therefore, considering only $\lambda>0$, it also works for the cases where $E=-2$, $\expval{\epsilon}<0$ and $E=2$,  $\expval{\epsilon}>0$.  However, when $|E|=2$, the expansion for $h(\lambda,E)$ is different as it now has  half integer powers of $\lambda$ also. This makes the leading contribution to the Lyapunov exponent,  $\zeta(\lambda)\sim \lambda$ for the two cases.

For $|-E+\expval{\epsilon}\lambda|\geq 2$, the above proof does not work. The cases where this happens are  $|E|=2$ with  $\expval{\epsilon}=0$, $E=2$ with  $\expval{\epsilon}<0$, and   $E=-2$ with $\expval{\epsilon}<0$. Let us first consider the latter two cases. For these cases the Lyapunov exponent is finite even if the disorder is replaced by its average, $\expval{\epsilon}$. In that case, the solution for Eq.~(\ref{eq:itereq}) is given by $p_L(\lambda)=\frac{\sin (L+1) h(\lambda)}{\sin h(\lambda)}.$  Therefore, in the limit $L\rightarrow\infty$, $p_L(\lambda)\sim e^{\sqrt{\expval{\epsilon} \lambda}L}$ which gives $\zeta(\lambda)\sim \lambda^{1/2}$. The disorder only contributes at higher orders in $\lambda$. The case where $\expval{\epsilon}=0$ is very subtle and requires an elaborate proof. We point the reader to Ref.~\onlinecite{Cane_2021}  which considers Eq.~(\ref{eq:itereq}) for $E=-2$ in the context of harmonic wires with disordered magnetic fields.  In this work, the Lyapunov exponents are determined by mapping Eq.~(\ref{eq:itereq}) for $E=-2$ to a harmonic oscillator with noisy frequency which is well studied in literature~\cite{crauel1999perturbation}. It is shown that  $\zeta(\lambda)\sim \lambda^{2/3}$ for $\expval{\epsilon}=0$, and also $\zeta(\lambda)\sim \lambda^{1/2}$ and $\zeta(\lambda)\sim \lambda$ for $\expval{\epsilon}>0$ and $\expval{\epsilon}<0$,   respectively.

	\section{Scaling of the Conductance}
	\label{sec:plaw}
	We now determine the scaling of the average behavior of the conductance. Let us assume $\zeta(\lambda_{\mathbf{k}})\sim |\lambda_{\mathbf{k}}|^a$, where $a$ is known from our previous results. Therefore, the $|\lambda_{\mathbf{k}}|$ values for which $|\lambda_{\mathbf{k}}|^a L<1$  contribute in the integral for the conductance in Eq.~(\ref{eq:condint1}). Hence, in the limit $L\rightarrow\infty$, we can cutoff the integral over $\mathbf k$ as follows,
	\begin{equation}
		\expval{\tau(E)}\approx\frac{4}{\pi}\lim\limits_{L\rightarrow\infty}\int_{\mathbf{k}_*-\mathbf{k}_c}^{\mathbf{k}_*+ \mathbf{k}_c} d\mathbf{k} \expval{|F_L(\lambda_{\mathbf{k}})|^2},\label{eq:cuttau}
	\end{equation}
	where $\mathbf{k}_*$ is the point such that $\lambda_{\mathbf{k}_*}=0$, and $\mathbf{k}_c>0$  is a small deviation from $\mathbf{k}_*$ such that $|\lambda_{\mathbf{k}_*+\mathbf{k}_c}|^aL\sim 1$. Let the Taylor expansion of $\lambda_{\mathbf{k}}$ around $\mathbf k=\mathbf{k}_*$ be $\lambda_{\mathbf{k}_*+\mathbf{k}_c}=\lambda_0 \mathbf{k}_c^b+\order{\mathbf{k}_c^{b+1}}$, we have $\mathbf{k}_c\sim 1/L^{1/(ab)}$. 

	Within the range of integration in Eq.~(\ref{eq:cuttau}), the disorder is effectively absent so we make an assumption by replacing $\expval{|F_L(\lambda_{\mathbf{k}})|^2}$ by $|F_L^o(\lambda_\mathbf k)|^2$ which is same quantity computed with the disorder replaced by its average at every $x$. With some simple algebra we can show,	
	\begin{equation}
		F_L^o(\lambda_{\mathbf{k}})=\frac{\gamma \sin q_{\mathbf k}}{\sin q_{\mathbf k}(L+1)+2i\gamma \sin q_{\mathbf k} L-\gamma^2\sin q_{\mathbf k} (L-1)}\label{eq:FLo1}.
	\end{equation}
where $q_\mathbf k=\arccos[(-E+\lambda_\mathbf k \expval{\epsilon})/2]$. 
Using this assumption Eq.~(\ref{eq:cuttau}) reduces to,
\begin{equation}
	\expval{\tau(E)}\approx\frac{4}{\pi}\lim\limits_{L\rightarrow\infty}\int_{\mathbf{k}_*-\mathbf{k}_c}^{\mathbf{k}_*+ \mathbf{k}_c} d\mathbf{k} |F_L^o(\lambda_{\mathbf{k}})|^2.\label{eq:cuttauapprox}
\end{equation}
	For $|E|<2$, the integrand in Eq.~(\ref{eq:cuttauapprox}) is finite at $\mathbf k =\mathbf k_*$ and is highly oscillatory with $L$ around $\mathbf k_*$. However, as $L\rightarrow\infty$,   it  approximately averages out to, 
	$
	|\bar{F}^o(q_{\mathbf k})|^2\approx \big|\frac{\gamma}{2(1+\gamma^2)}\sin q_{\mathbf k}\big|^2,
	$
	under the integral sign
	 (see  Appendix \ref{app:cond2}). Using $\bar{F}^o(q_\mathbf k)$ in Eq.~(\ref{eq:cuttauapprox}), we get $\expval{\tau(E)}\sim 1/L^{ab}$. Now,  $a=2$ for all values of $|\mu|\leq 2$ but  $b=1$ for $|\mu|<2$ and $b=2$ for $|\mu|=2$ . Therefore, we expect  $1/L^{1/2}$ and $1/L^{1/4}$ for these two cases, respectively, and the numerical results shown in panels A1 and A2 of Fig.~(\ref{fig:powlaw1}) are in good agreement.  For $|E|>2$ or  $|\mu|>2$ the integrand decays exponentially so does the conductance.

For $|E|=2$ the integrand requires a bit more attention as $q_{\mathbf{k_*}}=\pi$ which means $F_L^o(\lambda_{\mathbf{k_*}})$  vanishes. Therefore, its behavior  around $\mathbf k_*$  matters in the scaling of the conductance.  This  behavior around $\mathbf k_*$ is different  for real and imaginary $q_\mathbf k$. 
 Let us first consider $q_\mathbf k$ to be real, then from Eq.~(\ref{eq:FLo1}),  $F_L^o(\lambda_{\mathbf{k}})$ is highly oscillatory with $L$ and once again  averages out to 
$
\bar{F}^o(q_{\mathbf k}) \sim |\mathbf{\bar{k}}|^{b/2},~\mathbf{\bar{k}}=\mathbf{k}-\mathbf{k_*}.
$
 Using this in Eq.~(\ref{eq:cuttauapprox}) we have,
$
	\expval{\tau(E)}\sim \int_{0}^{L^{1/(ab)}}d\mathbf{ \bar{k}}~~\mathbf{\bar k}^{b/2}=1/L^{\frac{b+2}{2 ab}}. \label{eq:cuttauapprox1}
$
 This power-law  is  valid only if $q_\mathbf k$  is real, equivalently $\abs{-E+\lambda_0\expval{\epsilon}\mathbf{\bar k}^b}<2,$ within the range of the integration. Therefore, the sign of the term $\lambda_0\expval{\epsilon}\mathbf{\bar k}^b$  is  important as  $|E|=2$ . Depending on sign of $\expval{\epsilon}$ and the corresponding  values of $b$  and $a$, we find several different cases. These cases have  $E$, $\mu$ values which correspond to $1/L$ marked  corners  of the shaded square   and its horizontal edges in Fig.(\ref{fig:chart}a) for $\expval{\epsilon}<0$   as well as in Fig.(\ref{fig:chart}c) for $\expval{\epsilon}>0$. At those corners $b=2$, $a=1$ giving a diffusive scaling $\expval{\tau(E)}\sim 1/L$, and at the horizontal edges $b=1$, $a=1$ giving $\expval{\tau(E)}\sim 1/L^{3/2}$, respectively. We show a comparison between numerical computations and the predicted power-laws in panels A3, A4 of Fig.~(\ref{fig:powlaw1}) and once again we see a good agreement.

 Let us now  consider $q_\mathbf k$ to be imaginary which means $|-E+\lambda_0\expval{\epsilon}\mathbf{\bar k}^b|\geq 2$. This happens for the cases which correspond to the horizontal edges (B1) and the corners (B2) of Fig.~(\ref{fig:chart}b) for $\expval{\epsilon}=0$ and to the $1/L^3$ marked corners (B3)  of Fig.~(\ref{fig:chart}a) and Fig.~(\ref{fig:chart}c). For B1 and B2, $\expval{\epsilon}=0$ and therefore, $\lim\limits_{L\rightarrow\infty}\lim\limits_{q_{\mathbf{k}_*}\rightarrow\pi}F_L^o(\lambda_{\mathbf{k}})\sim 1/L$. For the case B3, $\expval{\epsilon}\neq 0$ but $\mathbf{k}_c\sim \frac{1}{L}$ as $a=1/2$ and $b=2$, and once again $\lim\limits_{L\rightarrow\infty}\lim\limits_{q_{\mathbf{k}_*}\rightarrow\pi}F_L^o(\lambda_{\mathbf{k}})\sim 1/L$~\cite{PhysRevB.109.125415,RaySub}. Using this behavior of $F_L^o(\lambda_{\mathbf{k}})$ and the corresponding  values of $a$ and $b$, we get $1/L^{7/2}$, $1/L^{11/4}$ and $1/L^{3}$ for B1, B2 and B3, respectively.

Panels B1, B2 and B3 in Fig.~(\ref{fig:powlaw1}) show a comparison with the numerically calculated conductance for these cases. We see  that the theoretical arguments underestimate the power-law approximately by a factor of $L$ for B1 and B2. While the theory predicts $1/L^{7/2}$ and $1/L^{11/4}$ for the two cases, numerical data agrees with $1/L^{5/2}$ and $1/L^{7/4}$, respectively. The reason being that the approximation that $|F_L^o(\lambda_{\mathbf{k}})|^2=\expval{|F_L(\lambda_{\mathbf{k}})|^2}$ for $\lambda_{\mathbf{k}}\rightarrow0$ fails, and in fact it underestimates $\expval{|F_L(\lambda_{\mathbf{k}})|^2}$.  This can be seen in panel X of Fig.~(\ref{fig:powlaw1}) where we plot a comparison between $|F_L^o(\lambda_{\mathbf{k}})|^2$ and $\expval{|F_L(\lambda_{\mathbf{k}})|^2}$ for the parameter values corresponding to the case B1. It is not clear how to estimate $\expval{|F_L(\lambda_{\mathbf{k}})|^2}$ and predict the observed power-laws  for these two cases, and therefore our work opens up an interesting question for further studies. A similar issue has also been pointed out for  harmonic wires in presence of disordered magnetic fields~\cite{Cane_2021}. For the case B3, we see a perfect agreement between the theory and the numerical computations.


\section{Conclusions}
\label{sec:concl}
In conclusion, we looked at a model of a disordered Fermionic wire in two-dimensions and studied the scaling of conductance with the length of the wire along the direction of transport. In particular, we find that despite the presence of uncorrelated disorder along the direction of the transport, the conductance shows a super-diffusive scaling. This is attributed to the presence of eigenstates with diverging localization lengths at energies with absolute values less than a cutoff $E_c$.  Using heuristic arguments, we determined the super-diffusive behavior to be $1/L^{1/2}$ which agrees with the numerical computations. We also showed that at $|E|=E_c$ and at some special values of the parameters of the wire the conductance shows various  sub-diffusive scalings and also diffusive scaling. The  sub-diffusive power-laws are sensitive to the the sign of the expectation value of the disorder and are also different if the expectation value of the disorder vanishes. 

Our heuristic arguments predict the different power-laws at $|E|=E_c$  except when the disorder average vanishes.  For this case, the assumptions underlying the heuristic arguments break-down and underestimate the numerically observed power-law by a factor of $L$. This case therefore requires further study in order to correctly predict the numerically observed power-laws.  

Finally, we comment on the possibility of experimental systems where our model  could be realized.  It seems difficult to find a material system with the hopping parameters of our model. However, with the developments in synthesizing arbitrary lattice models using degenerate cavity systems~\cite{wang2019synthesizing}, it could be possible to realize our proposed Hamiltonian and observe its rich transport behaviors. In fact, such optical cavities have been already used to simulate 2D topological insulators, and their edge-state transport properties~\cite{luo2015quantum}.

\section{Acknowledgments}
	I thank Marko \v{Z}nidari\v{c} and Yu-Peng Wang for useful discussions.  I also acknowledge support by Grant No. J1-4385 from the Slovenian Research Agency.
	
\bibliography{biblio}
\appendix
	\section{The Reservoir Hamiltonians}
	\label{app:res_hamil}
	Let us label the  annihilation and creation operators at a site $(x,y)$ as ($\psi(x,y)$, $\psi^\dagger(x,y)$), and $(\phi_{L/R}(x,y), \phi^\dagger_{L/R}(x,y))$ on the wire, the left lead and the right lead, respectively. These satisfy the usual Fermionic anti-commutation relations. Free boundary conditions are imposed at the horizontal edges of the reservoir and the system at $y=1$ and $y=W$, respectively. The contacts between the wire and the reservoirs are themselves modeled as tight-binding Hamiltonians, $\mathcal{H}_{WL}$ and $\mathcal{H}_{WR}$. To write the full Hamiltonian of system, we define  column vectors $\Psi(x)$, and $\Phi_{L/R}(x)$ of $W$ components  with the $y^{th}$ component given by the operators $\psi(x,y)$ and $\phi_{L/R}(x,y)$, respectively. Therefore, we have the following for the full Hamiltonian of the system,
	\begin{equation}
		\mathcal{H}=\mathcal{H}_L+ \mathcal{H}_{LW}+ \mathcal{H}_W+ \mathcal{H}_{RW}+\mathcal{H}_R,
	\end{equation}
	where the individual Hamiltonians of the wire, the contacts, and the reservoirs  are given by,
	\begin{widetext}
		\begin{align}
			\label{eq:wirehamil}\mathcal{H}_W&=\sum_{x=1}^{L}\epsilon_x \Psi^\dagger(x)H_0\Psi(x)+  \sum_{x=1}^{L-1}\left[\Psi^\dagger(x)\Psi(x+1)+ \text{h.c.}\right],\\
			\mathcal{H}_{LW}&= \eta_c (\Psi^\dagger(1)\Phi_L(0)+\Phi_L^\dagger(0)\Psi(1)),\\
			\mathcal{H}_{RW}&= \eta_c (\Psi^\dagger(L)\Phi_R(L+1)+\Phi_R^\dagger(L+1)\Psi(L)),\\
			\mathcal{H}_{L}&=\sum_{x=-\infty}^{0} \Phi^\dagger_L(x)H_{0}'\Phi_L(x)+\eta_{bx} \sum_{x=-\infty}^{-1}\Phi_L^\dagger(x)\Phi_L(x+1)+ \Phi_L^\dagger(x+1)\Phi_L(x),\\
			\mathcal{H}_{R}&=\sum^{\infty}_{x=L+1} \Phi^\dagger_R(x) H_{0}'\Phi_R(x)+\eta_{bx} \sum^{\infty}_{x=L+1}\Phi_R^\dagger(x)\Phi_R(x+1)+ \Phi_R^\dagger(x+1)\Phi_R(x).
		\end{align}
	\end{widetext}
	Note that we have taken the reservoir Hamiltonians to be the same and the couplings at the contacts are assumed to be of strength $\eta_c$.
	\section{Simplification of NEGF Conductance}
	\label{app:cond1}
	The NEGF formula for conductance for arbitrary lattice models is given by,
	\begin{equation}
		T(E)=4\pi^2 \Tr[ G^{+}(E) \Gamma_L(E)  G^-(E) \Gamma_R(E)],\label{eqs:cond1}
	\end{equation}
	where $G^+(E)=(E-H_W-\Sigma_L(E)-\Sigma_R(E))^{-1}$ is the effective Green's function of the wire and $\Gamma_{L/R}=(\Sigma_{L/R}^\dagger-\Sigma_{L/R})/(2\pi i)$.
	Since the contacts with the reservoirs are only along the edges i.e. at $x=1$, and $x=L$, the  trace in Eq.~(\ref{eqs:cond1}) can be computed as,
	\begin{equation}
		T(E)=4\pi^2 \Tr[ G^{+}_{1L} \Gamma  G^-_{L1} \Gamma],
		\label{eq:cond_2}
	\end{equation}
	where $G^+_{1L}$ is a $W\times W$ matrix with components given by,  $G^+_{1L}[y,y']=G^+(E)[x=1,y;x'=L,y']$ and $G^{-}_{1L}=[G^+_{1L}]^\dagger$. $\Gamma$ is the only non-zero block of $\Gamma_{L/R}$ given by, $\Gamma[y,y']=\Gamma_R[L,y;L,y']=\Gamma_L[1,y;1,y']$.

	Using the transfer matrix approach, it can be shown that~\cite{PhysRevB.109.125415},
	\begin{equation} 
		\tilde G^+_{1L}=[P_{L}+\tilde \Sigma P_{L-1}+Q_{L}\tilde \Sigma +\tilde \Sigma Q_{L-1}\tilde \Sigma ]^{-1}
	\end{equation}
	The tilde over the Green function and self energy matrices denotes that these are written in the diagonal basis of $H_0$ i.e. $\tilde G^+_{1L}[y,y']=- U G^+_{1L} U^{\dagger}$.  Similarly, $\tilde\Sigma[y,y']=U\Sigma[y,y']U^\dagger=U\Sigma_R[L,y;L,y']U^\dagger=U\Sigma_L[1,y;1,y']U^\dagger$, is the only non-zero block of the matrices $\Sigma_{L}$ and $\Sigma_{R}$. $P_L$ and $Q_L$ are diagonal matrices with entries given by $p_L(\lambda_k)$ and $q_L(\lambda_k)$, $k=1,2,3,...,W$, which are defined in the main text.
	
	It can be shown that $\Sigma=U_L \Sigma_D  U_L^\dagger$, where $\Sigma_D$ is a diagonal matrix  with  components given by~\cite{dhar2006},
	
	\begin{align}
		\Sigma_D[k,k]=\begin{cases}
			\frac{\eta_c^2}{\eta_{bx}}[z_k+i \sqrt{1-z_k^2}]; & |z_k|<1,\\
			\frac{\eta_c^2}{\eta_{bx}}[z_k- \sqrt{z_k^2-1}]; & z_k>1,\\
			\frac{\eta_c^2}{\eta_{bx}}[z_k+ \sqrt{z_k^2-1}]; & z_k<-1,
		\end{cases}
		\label{eq:sed}
	\end{align}
	where $z_k=\frac{E- \lambda_k'}{2\eta_{bx}}$. $U_L$ is the unitary transformation that diagonalizes the intra-chain hopping matrix  $H'_0$ of the reservoirs and its eigenvalues are given by $ \lambda_k'$. We consider the limit where $|\eta_{bx}|>>|\omega-\tilde\lambda_{k}|$ for all $k$. Note that if $H_0'$ is chosen to correspond to  nearest-neighbor hopping model with hopping $\eta_{by}$, then $\lambda_k'=2\eta_{by}\cos[\frac{k\pi}{W+1}]$ and the limiting condition reads $|\eta_{bx}|>>|E|+2|\eta_{by}|$. In this limit, $\tilde\Sigma\approx\Sigma\approx\Sigma_D\approx i \gamma I$ and $\Gamma\approx \frac{\gamma}{\pi}I$, where $\gamma=\eta_c^2/\eta_{bx}$. Thus, Eq.~(\ref{eq:cond_2}) reduces to,
	\begin{align}
		&T(E)\notag\approx\\&4\sum_{k=1}^W \frac{\gamma^2}{|p_{L}(\lambda_{k})+i \gamma [p_{L-1}(\lambda_{k})+q_{L}(\lambda_{k})]- \gamma^2 q_{L-1}(\lambda_{k})|^2}\\&=4\sum_{k=1}^W |F_L(\lambda_{k})|^2.\label{eqs:cond4}
	\end{align}
	The sum over $k$ in Eq.~(\ref{eqs:cond4})  sums over the eigenvalues of $H_0$ and  in the limit $W\rightarrow\infty$, this sum  can be replaced by an integral over the spectrum of $H_0$ given by $\mu+2\cos(\mathbf{k}),~\mathbf{k}\in(0,\pi)$,
	\begin{align}
		&\tau(E)=\frac{T(E)}{W}\approx\frac{4}{\pi}\int_0^\pi d\mathbf k |F_L(\lambda_\mathbf k)|^2,\label{eq:condint}
	\end{align}
	where $\tau(E)$ is the conductance per unit width of the wire.
	\section{ Lyapunov Exponent Proof}
	\label{appen:lyap}
	We begin rewriting Eq.~(\ref{eq:zetatheta}) as,
	\begin{align}
		\zeta(\lambda)&=\int_{-\pi/2}^{\pi/2}d\theta ~\mathbf P[\theta,\lambda] \log\abs{ \frac{\cos(\theta+h(\lambda,E))}{\cos\theta}}\\
		&=\int_{-\pi/2}^{\pi/2}d\theta ~(\mathbf P[\theta-h(\lambda),\lambda]-\mathbf P[\theta,\lambda]) \log|\cos\theta|\label{eq:zetatheta1}
	\end{align}
	We have used the fact that $\mathbf P[\theta,\lambda]$ is periodic as the iteration equation, Eq.~(\ref{eq:itertheta}), is itself periodic in the variable $\theta$. We Taylor expand Eq.~(\ref{eq:zetatheta1}) around $\lambda=0$, to obtain the coefficients $\boldsymbol\zeta_0(\theta,E)$, $\boldsymbol{\zeta}_1(\theta,E)$ and $\boldsymbol{\zeta}_2(\theta,E)$ as, 

		\begin{align}
			\boldsymbol{\zeta}_0(\theta,E)&= \mathbf{P}_0(\theta-h_0(E))-\mathbf P_0(\theta)\\
			\boldsymbol{\zeta}_1(\theta,E)&=-h_1(E) \partial_\theta\mathbf P_0(\theta-h_0(E))+\mathbf P_1(\theta-h_0(E))-\mathbf P_1(\theta)\\	
			\boldsymbol{\zeta}_2(\theta,E)&=-h_2(E)\partial_\theta\mathbf P_0(\theta-h_0(E)) +\frac{h_1(E)^2}{2}\partial_\theta^2 \mathbf P_0(\theta-h_0(E)\notag\\&-h_1(E)\partial_\theta \mathbf  P_1(\theta-h_0(E))+\mathbf P_2(\theta-h_0(E))-\mathbf{P}_2(\theta).\label{eq:z2} 
		\end{align}
		Let us now Taylor expand the self consistency equation of $\mathbf P[\theta,\lambda]$, Eq.~(\ref{eq:selfP}). For the zeroth, first and second order we  have,
	\begin{widetext}	
		\begin{align}
			&\mathbf{P}_0(\theta)-\mathbf P_0(\theta-h_0(E))=0\label{eq:P0},\\&\label{eq:P1} \mathbf P_1(\theta)-\bigg[\partial_\theta \mathbf P_{0}(\theta-h_0(E))\expval{\Theta_1}+\mathbf P_1(\theta-h_0(E))+\partial_\theta \expval{\Theta_1} \mathbf P_0(\theta-h_0(E))\bigg]=0,\\&\label{eq:P2}\mathbf P_2(\theta)-\bigg[\mathbf P_2(\theta-h_0(E)) +\partial_\theta \mathbf P_0(\theta-h_0(E))\expval{\Theta_2} +\frac{1}{2}\partial_\theta^2\mathbf P_0(\theta-h_0(E)) \expval{\Theta_1^2}+\partial_\theta \mathbf P_1(\theta-h_0(E)) \expval{\Theta_1}\notag\\&+\mathbf P_0(\theta-h_0(E)) \partial_\theta \expval{\Theta_2}+(\partial_\theta \mathbf P_{0}(\theta-h_0(E))\expval{\Theta_1}+\mathbf P_1)(\partial_\theta \expval{\Theta_1})\bigg]=0,
		\end{align}
		where we assumed $\Theta[\theta,\lambda]=\theta-h_0(E)+\sum_{i=1}^\infty \Theta_{i}(\theta)\lambda^{i}$.
	\end{widetext}
	Eq.~(\ref{eq:P0}) immediately gives $\boldsymbol{\zeta}_0(\theta,E)=0$. To show that $\boldsymbol{\zeta}_1(\theta,E)=0$, we consider the expansion of functions $\expval{\Theta[\theta,\lambda]}$ and $h(\lambda,E)$  around $\lambda=0$,
	\begin{align}
		h(\lambda)&=h_0(E)-\frac{\expval{\epsilon}\lambda}{(4-E^2)^{1/2}}+\frac{E\expval{\epsilon}^2\lambda^2}{2(4-E^2)^{3/2}}+\order{\lambda^3} 
	\end{align}
\begin{align}
		\expval{\Theta[\theta,\lambda]}&=\theta-h_0(E)+\frac{\expval{\epsilon}\lambda}{(4-E^2)^{1/2}}\notag\\&-\left[\frac{E\expval{\epsilon}^2+8\sqrt{4-E^2}\sigma^2\cos^3\theta\sin\theta}{2(4-E^2)^{3/2}}\right]\lambda^2+\order{\lambda^3}\label{eq:Thetaexp}
	\end{align}
	where $\sigma^2=\expval{(\epsilon_x-\expval{\epsilon})^2}$ and $h_0(E)=\arccos[-E/2]$. Note that $\expval{\Theta_1}=-h_1(E)$ and $\partial_\theta\expval{\Theta_1}=0$. Using these two in  Eq.~(\ref{eq:P1})  we get,
	\begin{align}
		\label{eq:P11} \mathbf P_1(\theta)+h_1(E)\partial_\theta \mathbf P_{0}(\theta-h_0(E))-\mathbf P_1(\theta-h_0(E))=0
	\end{align}
	and therefore $\boldsymbol{\zeta}_1(\theta,E)=0$. Let us now compute $\boldsymbol{\zeta}_2$, for this note that Eq.~(\ref{eq:P0}) implies that $\mathbf{P}_0(\theta)$ is a constant, as the equation holds for arbitrary $|E|<2$. Fixing  $\mathbf P_0(\theta)$ using normalization to $\frac{1}{\pi}$, and substituting in Eq.~(\ref{eq:z2}) and Eq.~(\ref{eq:P2}) we get,
	\begin{align}
		&\boldsymbol{\zeta}_2(\theta,E)=-h_1(E)\partial_\theta \mathbf  P_1(\theta-h_0(E))\notag\\&~~~~~~~~~~~~~~~~+\mathbf P_2(\theta-h_0(E))-\mathbf{P}_2(\theta),\label{eq:z21}\\
		&-\mathbf P_2(\theta)+\mathbf P_2(\theta-h_0(E))\notag\\&~~~-h_1(E) \partial_\theta \mathbf P_1(\theta-h_0(E))+\frac{1}{\pi}\partial_\theta \expval{\Theta_2}=0
	\end{align}  
	Comparing these two equations, we get $\boldsymbol{\zeta}_2(\theta,E)=-\frac{1}{\pi}\partial_\theta \expval{\Theta_2}$ where $\Theta_2$ is given by Eq.~(\ref{eq:Thetaexp}). Therefore, the Lyapunov exponent is given by,
	\begin{align}
		\zeta(\lambda)&=-\frac{\lambda^2}{\pi} \int_{-\pi/2}^{\pi/2}d\theta \log |\cos\theta| \partial_\theta\expval{\Theta_2}+\order{\lambda^3}\\&= \frac{\sigma^2}{2(4-E^2)}\lambda^2+\order{\lambda^3}.
	\end{align}
	Note that the pre-factor of $\lambda^2$ is positive only in the domain of the applicability of the solution, $|E|<2$. At $|E|=2$ it diverges, reminiscent of the fact that at this point the behavior is different.

	\section{Derivation of $\bar F^o(q_{\mathbf k})$}
	\label{app:cond2}

	If $\epsilon_x=\expval{\epsilon}$ for all $x$, then it is straightforward to see that $p_L(\lambda_{\mathbf{k}})=\sin[q_\mathbf k(L+1)]/\sin[q_\mathbf k]$, which gives
	\begin{equation}
		F_L^o(\lambda_{\mathbf{k}})=\frac{\gamma \sin q_{\mathbf k}}{\sin q_{\mathbf k}(L+1)+2i\gamma \sin q_{\mathbf k} L-\gamma^2\sin q_{\mathbf k} (L-1)}\label{eq:FLo}.
	\end{equation}
	We rewrite this expression for $F_L^o(\lambda_{\mathbf{k}})$ as follows,
	\begin{equation}
		F_L^o(\lambda_{\mathbf{k}})=\frac{\gamma\sin q_{\mathbf k}}{A(q_{\mathbf k}) \sin L q_{\mathbf k}+ B(q_{\mathbf k})\sin q_{\mathbf k} \cos L q_{\mathbf k}}, 
	\end{equation}
	where $A(q_{\mathbf k})=(1-\gamma^2)\cos q_{\mathbf k}+ 2 i \gamma$ and $B(q_{\mathbf k})=(1+\gamma^2)$.
	Therefore, for a disorder free chain we have,
	\begin{align}
		&\tau_0(E)=\frac{4\gamma^2}{\pi}\int_{0}^{\pi} d \mathbf k\frac{\sin^2 q_{\mathbf k}}{|A(q_{\mathbf k}) \sin L q_{\mathbf k}+ B(q_{\mathbf k})\sin q_{\mathbf k} \cos L q_{\mathbf k}|^2}\\&=\frac{8\gamma^2}{\pi}\int_{0}^{\pi} d \mathbf k\sin^2 q_{\mathbf k}\frac{[|A(q_{\mathbf k})|^2+|B(q_{\mathbf k})|^2\sin^2 q_{\mathbf k}]^{-1}}{1+R(q_{\mathbf k}) \sin[2 L q_{\mathbf k} +\theta(q_{\mathbf k})]}\label{eqs:keq}
	\end{align}
	\begin{align}	
		\approx \frac{8\gamma^2}{\pi}\int_{0}^{\pi}dq \abs{\frac{d \mathbf k}{d q}} \sin^2 q\frac{[|A(q)|^2+|B(q)|^2\sin^2 q]^{-1}}{1+R(q) \sin[2 L q +\theta(q)]}\label{eqs:qkeq},
	\end{align}	
	where, $R(q_{\mathbf k})$ and $\theta(q_{\mathbf k})$ are defined via the relations,
	\begin{align}
		R(q_{\mathbf k})\cos[\theta(q_{\mathbf k})]&= \frac{2\Re[A(q_{\mathbf k}) B^*(q_{\mathbf k})]\sin q_{\mathbf k}}{|A(q_{\mathbf k})|^2+|B(q_{\mathbf k})\sin q_{\mathbf k}|^2},\\
		R(q_{\mathbf k})\sin[\theta(q_{\mathbf k})]&= \frac{|A(q_{\mathbf k})|^2-|B(q_{\mathbf k})\sin q_{\mathbf k}|^2}{|A(q_{\mathbf k})|^2+|B(q_{\mathbf k})\sin q_{\mathbf k}|^2}.
	\end{align}
	Note while changing the integration variable from $\mathbf k$ to $q_\mathbf k$, we kept the limits of the integration same. This approximation loses some irrelevant details of $\tau_0(E)$ but helps to take the limit $L\rightarrow\infty$. We will discuss what details are lost when we present some numerical results by the end of this section. In the limit $L\rightarrow\infty$, we have, 
	\begin{align}
		\bar \tau_0(E)&=\frac{8\gamma^2}{\pi}\lim\limits_{L\rightarrow\infty}\int_{0}^{\pi}dq \abs{\frac{d \mathbf k}{d q}} \frac{\sin^2 q}{1+R(q) \sin[2 L q +\theta(q)]}\\
		&= \frac{8\gamma^2}{\pi}\int_{0}^{\pi}dq \abs{\frac{d \mathbf k}{d q}}\frac{\sin^2 q}{\sqrt{1-R^2(q)}}\label{eq:feq},
	\end{align}
	where the last step follows from the identity~\cite{roy2008heat},
	
	\begin{equation}
		\lim\limits_{N\rightarrow \infty} \int_0^\pi du \frac{g(u)}{1+h(u)\sin[2 u N+r(u)]}=\int_0^\pi du \frac{g(u)}{\sqrt{1-h^2(u)}}.
	\end{equation}
	Converting the integral in Eq.~(\ref{eq:feq}) back from $q$ to $k$ and then substituting $\frac{1}{\sqrt{1-R^2(q_{\mathbf k})}}=\frac{|A(q_{\mathbf k})|^2+|B(q_{\mathbf k})\sin q_{\mathbf k}|^2}{2 \Im[A(q_{\mathbf k})B^*(q_{\mathbf k})]\sin q_{\mathbf k}}$ we have,
	\begin{align}
		\bar \tau_0(E)&=\frac{8\gamma^2}{\pi}\int_{0}^{\pi} d\mathbf{k}  \frac{\sin q_{\mathbf k}}{2 \Im[A(q_{\mathbf k})B^*(q_{\mathbf k})]}\\&=\frac{4}{\pi}\int_{0}^{\pi} d\mathbf{k} \bar F^o( q_{\mathbf k} ).
	\end{align}
		\begin{figure}
		\subfigure[$E=\mu=0$]{\includegraphics[width=0.23\textwidth,page=3]{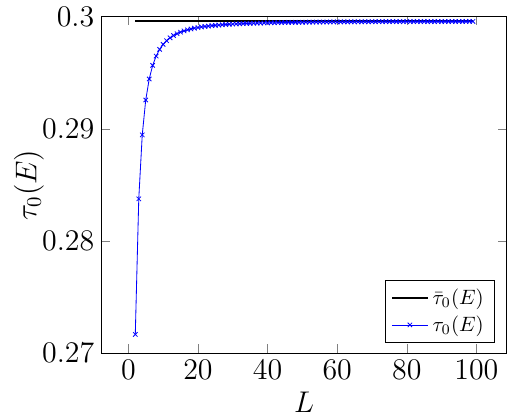}}
		\subfigure[$E=0.5,~\mu=0.3$]{\includegraphics[width=0.23\textwidth,page=4]{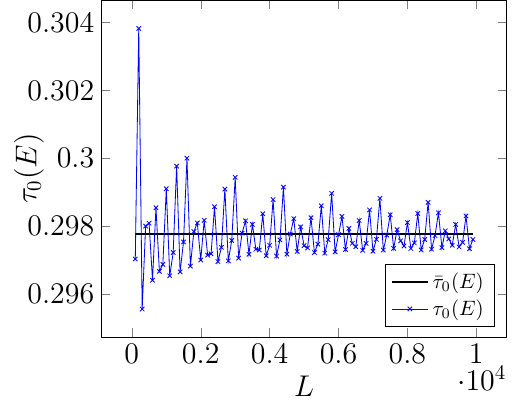}}
		\caption{Convergence of $\tau_0(E)$ to $\bar\tau_0(E)$. Parameter values: $W=10^4$, $\expval{\epsilon}=1$,  and $\gamma=0.25$.}	
		\label{fig:conv}
	\end{figure}
	We show in Fig.~(\ref{fig:conv}) the convergence of $\tau_0(E)$ to $\bar{\tau}_0(E)$ as $L$ is increased. Note that Fig.~(\ref{fig:conv}a), the parameters are such that $q_\mathbf k= \mathbf k$, so the change of limits from Eq.~(\ref{eqs:keq}) to Eq.~(\ref{eqs:qkeq}) is exact and we see a smooth convergence to the expected value from the theory. When the parameter values are such that the change of limits is not exact~Fig.~(\ref{fig:conv}b), we see highly oscillating behavior around the theoretically expected value in the numerical calculation. Though the oscillations are not captured by the approximation, they are irrelevant for the scaling of the conductance.

\end{document}